\documentclass[twocolumn,showpacs,preprintnumbers,amsmath,amssymb]{revtex4}

\input psfig.sty

\usepackage{graphicx}
\usepackage{dcolumn}
\usepackage{bm}

\begin{document}

\def\be{\begin{equation}}
\def\ee{\end{equation}}
\def\lb{\label}


\title{Nonextensive effects on the phase structure of QHD} 


\author{F. I. M. Pereira$^{1}$} \email{flavio@on.br}

\author{R. Silva$^{2,3}$} \email{raimundosilva@dfte.ufrn.br}

\author{J. S. Alcaniz$^{1}$} \email{alcaniz@on.br}

\affiliation{$^{1}$Observat\'orio Nacional, Rua Gal. Jos\'e Cristino 77, 20921-400, Rio de Janeiro RJ, Brasil}

\affiliation{$^{2}$Universidade Federal do Rio Grande do Norte, UFRN, Departamento de F\'{\i}sica C. P. 1641, Natal-RN, 59072-970, Brazil}

\affiliation{$^{3}$Universidade do Estado do Rio Grande do Norte, Departamento de F\'{\i}sica - UERN 59610-210, Mossor\'o - RN, Brasil}

\date{\today}

\begin{abstract}

We investigate nonextensive effects on phase transition in nuclear matter in the context Walecka many-body field theory. Changes in the general behaviour are observed when the results calculated for the nuclear matter at vanishing baryon density are compared to those obtained through the standard Fermi-Dirac distribution. It is observed a dependence between the nonextensive parameter $q$ and the coupling constants $C_S^2$ of the phase transition. A numerical relation for this thermodynamical dependence is also proposed.

\end{abstract}

\maketitle

\section{Introduction}

Non-extensive statistical mechanics (NESM) is a generalization of thermodynamics and statistical mechanics aiming to overcome a number of
physical systems that possess exotic properties, such as broken ergodicity, strong correlation between elements, multi fractality of
phase-space and long-range interactions. The NESM framework proposed by Tsallis \cite{TG} is based on the nonadditive $q$-entropy
\begin{equation}\label{1}
S_q=k{1-\sum_{i=1}^W p_i^q\over q-1},
\end{equation}
where $k$ is a positive constant, $W$ is the number of microscopic states, and $p_i$ is a normalized probability distribution. 

In fact, additivity for two probabilistically independent subsystems A and B is generalized by the following pseudoadditivity: 
$$
\frac{S_q (A , B)}{k} = \frac{S_q (A)}{k} + \frac{S_q (B)}{k} + (1 - q) \frac{S_q (A) S_q (B)}{k^2}\;.
$$ 
For subsystems that have special probability correlations, extensivity is not valid for Boltzmann-Gibbs (BG) entropy, but may occur for $S_q$ with a particular value of the index $q\neq 1$, called the $q$-entropic parameter. In the limit $q \rightarrow 1$, BG entropy is additive, i.e, 
$$
S_{BG}(A,B)=S_{BG} (A)+S_{BG}(B)\;,
$$ 
and Eq. (\ref{1}) reduces to the usual Boltzmann and Gibbs formula
$S_1=k\sum_{i=1}^W p_i\ln p_i$,
which is the fundamental expression describing systems that usually do not present the exotic properties described above.

Several consequences of this generalized framework have been investigated in the literature \cite{q-application} (see also \cite{update} for a regularly updated bibliography). This includes systems of interest in high energy physics, e.g., the problem of solar neutrino \cite{34,35}, the charm quark dynamics in a thermal quark-gluon plasma for the case of collisional equilibration \cite{36}, interpretations for central Au-Au collisions at RHIC energies in a Relativistic Diffusion Model (RDM) \cite{37}, among others \cite{new,new1}.

Recently, an interesting connection between quantum statistics and Tsallis NESM has been presented in Ref.~\cite{TPM}. The main result of this study is that, for $q>1$, a $q$-generalized quantum distributions for fermions and bosons are given by
\begin{equation}\label{nq}
 n_q(\mu,T)=\frac{1}{\tilde{e}_q(\beta(\epsilon-\mu))\pm1},
  \end{equation}
where $\tilde{e}_q$ reads
\begin{eqnarray}
\label{TPM}\tilde{e}_q(x)= \left\{
\begin{array}{l}
~[1+(q-1)x]^{\frac{1}{q-1}}~~~~~{\rm if}~~~~~~x>0~~~~~~~~~~~~\\
\\
~[1+(1-q)x]^{\frac{1}{1-q}}~~~~~{\rm if}~~~~~~x\leq0~.~~~~~~~~~\\
\end{array}
\right.\
\end{eqnarray}
and $x=\beta(\epsilon-\mu)$. In the $q\rightarrow1$ limit, the standard Fermi-Dirac distribution, $n(\mu,T)$, is recovered. As physically expected,  as $T\rightarrow0$, $n_q(\mu,T)\rightarrow n(\mu,T)$. This amounts to saying that for studies of the interior of neutron stars (where, in nuclear scale, $T \simeq 0$) we do not expect any nonextensive signature. On the other hand, in heavy ions collision experiments or in the interior of protoneutron stars, with typical stellar temperatures of several tens of MeV (1 MeV$=1.1065\times10^{10}$ K), nonextensive effects may appear. 

\begin{figure*}[t]
\centerline{\psfig{figure=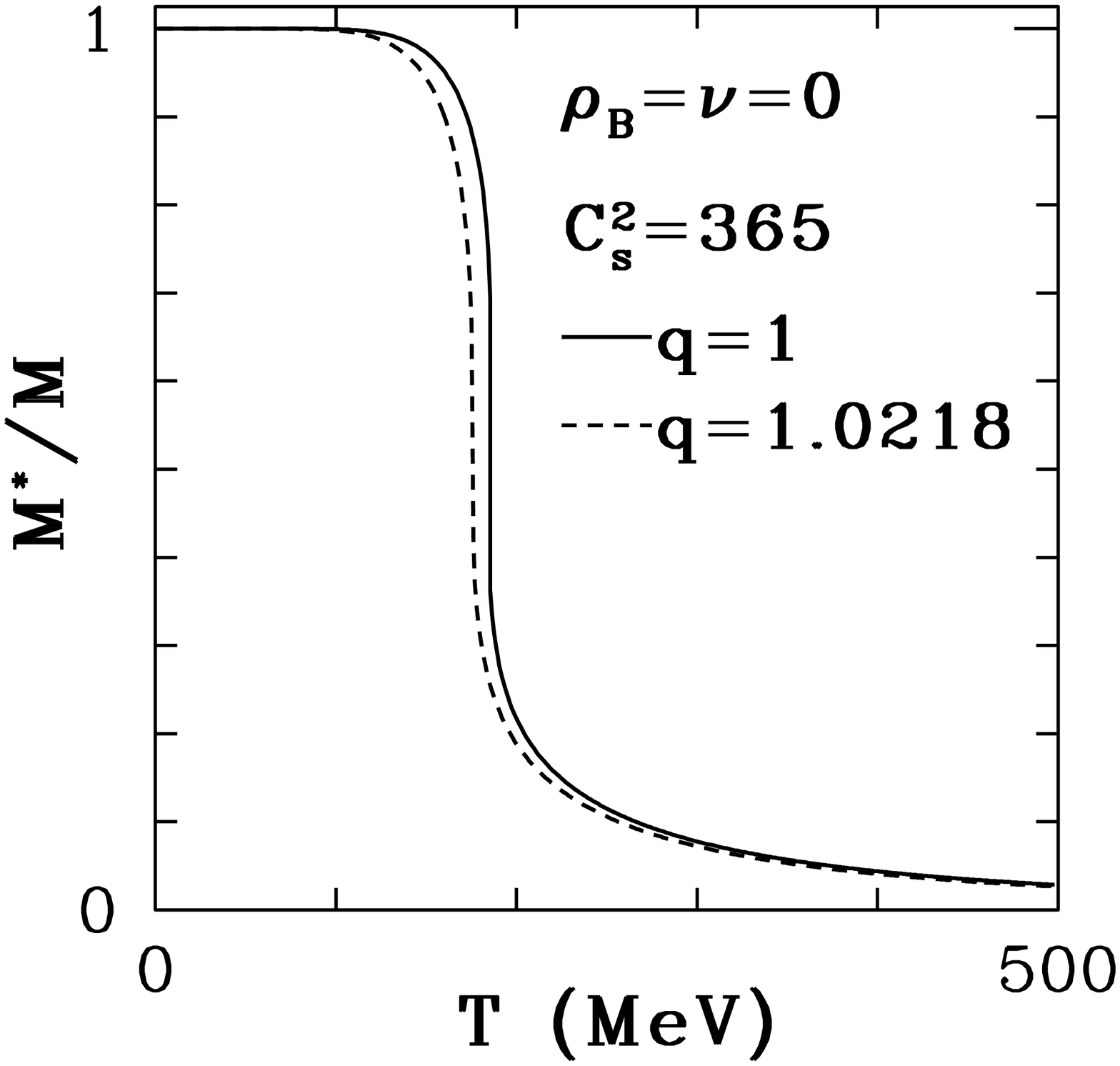,width=3.2truein,height=2.2truein}
\psfig{figure=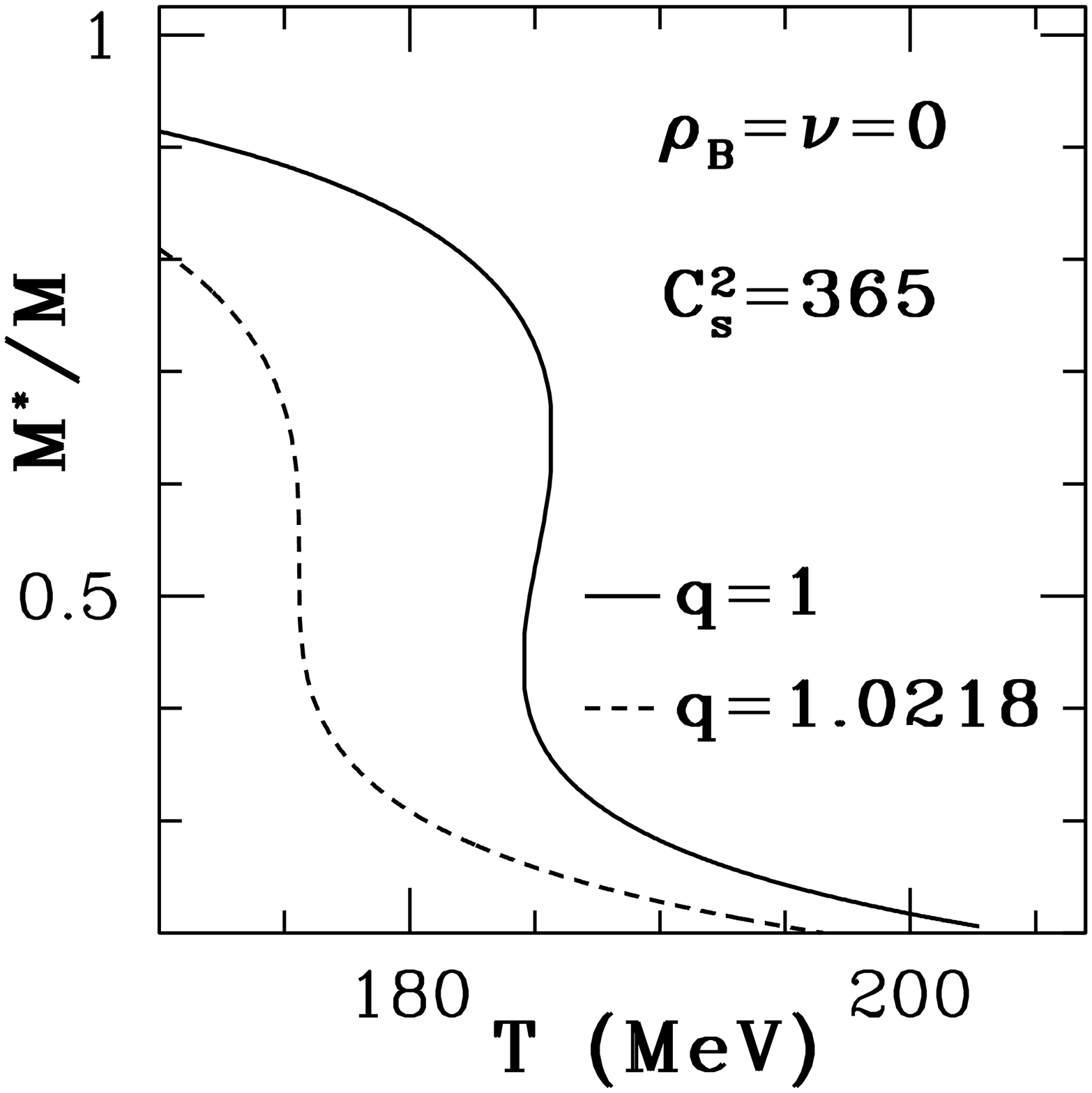,width=3.2truein,height=2.2truein}\hskip .5in}
\caption{The solution $M^*/M$ of Eq. (\ref{ms}) for nuclear matter ($\gamma=4)$ at vanishing baryon density as a function of temperature for the same value of $C_S^2$ and two different values of the parameter $q$. The right panel shows the same results but with a stretched temperature region around the point of phase transition.}
\label{fig1}
\end{figure*}

In this paper, by following the arguments of Ref. \cite{fimPRC2007}, we discuss non-extensive effects on nuclear matter at zero baryon density.  To this end we adopt  the relativistic phenomenological approach developed by Walecka~\cite{SW}, the so-called quantum hadrodynamics (QHD). By taking into account the results on the phase transion in nuclear matter discussed in Ref.~\cite{TSP},  we show that there exist a thermodynamical dependence between the statistical correlations characterized by non-extensive parameter $q$ and the coupling constant $C^2_S$. A numerical relation in the $q\times C^2_S$ plane resulting in regions of differents thermodynamical behaviors is also proposed.

\section{Nonextensivity on QHD}

In the mean-field approach (for details see reference \cite{SW}), we have the scalar density
\begin{equation}
\varrho_S=\frac{\gamma_{\rm
N}}{(2\pi)^3}\int\frac{M^*}{E^*(k)}[n_q(\nu,T)+n_q(-\nu,T)]d^{3}k~,
\label{RS}
\end{equation}
where $E^*(k)=\sqrt{k^2+{M^*}^2}$, $\beta=1/k_BT$, $M^*$ is the effective mass
\begin{equation}
M^*=M - g_{\sigma}\sigma = M-\frac{g_\sigma^2}{m_\sigma^2}\rho_S\;,
\label{ms}
\end{equation}
and, instead of the usual Fermi-Dirac distribution, we have used for $q>1$ the $q$-generalized quantum distribution for fermions,
$n_q(\nu,T)$, given by Eq. (\ref{nq}).

The baryon number density, the energy density and pressure are given,
respectively, by
\begin{equation}\label{RB}
\varrho_B=\frac{\gamma_{\rm
N}}{(2\pi)^3}\int[n_q(\nu,T)-n_q(-\nu,T)]d^{3}k,
\end{equation}
\begin{eqnarray}\label{edens}
\varepsilon&=&\frac{1}{2}\frac{g_\omega^2}{m_\omega^2}\varrho_B^2
+\frac{1}{2}\frac{g_\sigma^2}{m_\sigma^2}(M-M^*)^2+\nonumber\\
&&\frac{\gamma_{\rm N}}{(2\pi)^3}\int{E^*(k)}
[n_q(\nu,T)+n_q(-\nu,T)]d^{3}k\;,
\end{eqnarray}
\begin{eqnarray}\label{press}
p&=&\frac{1}{2}\frac{g_\omega^2}{m_\omega^2}\varrho_B^2
-\frac{1}{2}\frac{g_\sigma^2}{m_\sigma^2}(M-M^*)^2+\nonumber\\
&&\frac{1}{3}\frac{\gamma_{\rm N}}{(2\pi)^3}\int\frac{k^{2}}{E^*(k)}
[n_q(\nu,T)+n_q(-\nu,T)]d^{3}k\;,
\end{eqnarray}
The parameter
$\nu\equiv\mu-g_\omega\omega_0=\mu-(g_\omega/m_\omega)^2\varrho_B$
is the effective chemical potential, and $\gamma_N$ is the
multiplicity factor ($\gamma_N=2$ for pure neutron matter and
$\gamma_N=4$ for nuclear matter).

\begin{figure*}
\centerline{\psfig{figure=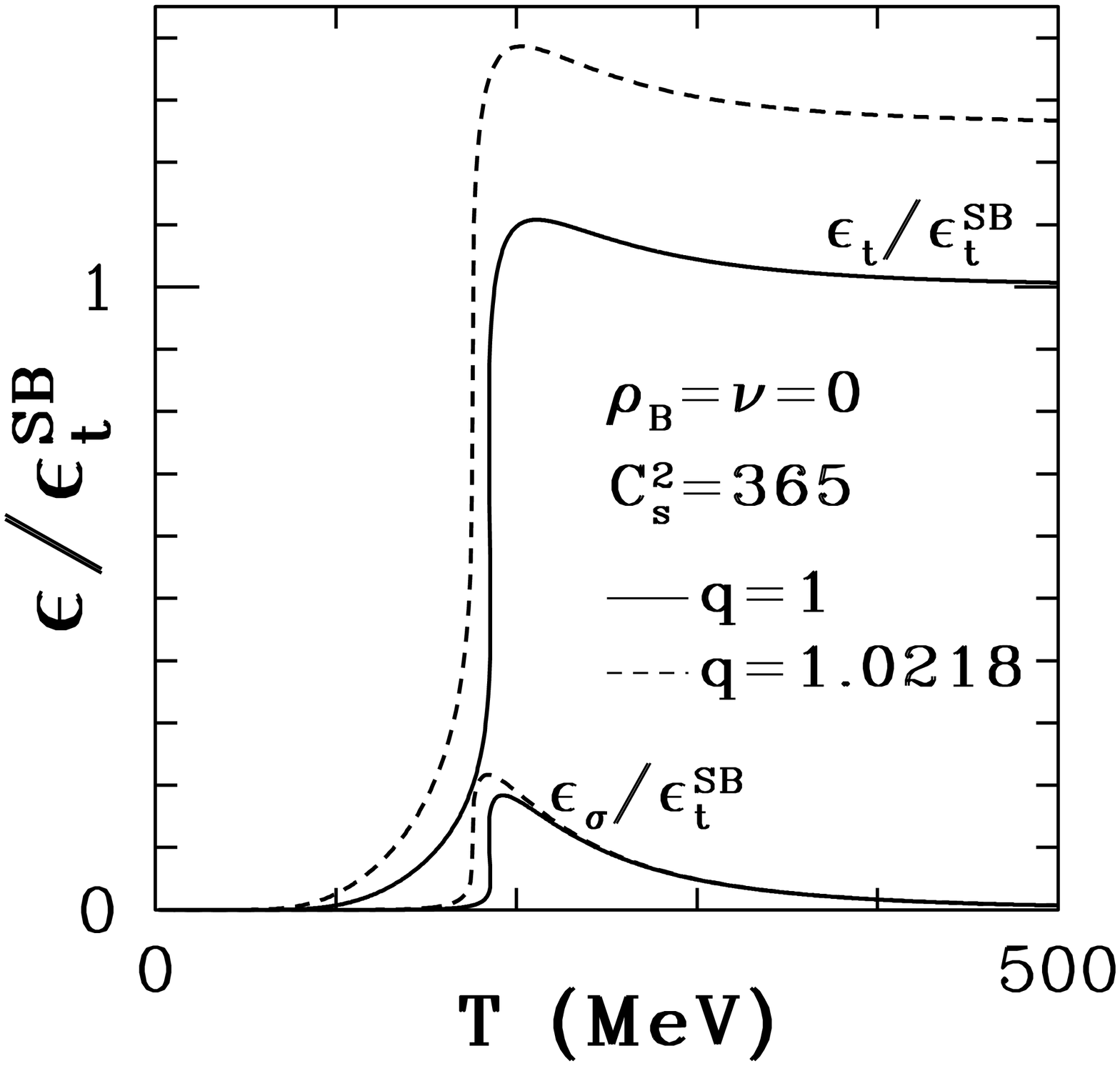,width=3.2truein,height=1.8truein}\hskip
.25in
\psfig{figure=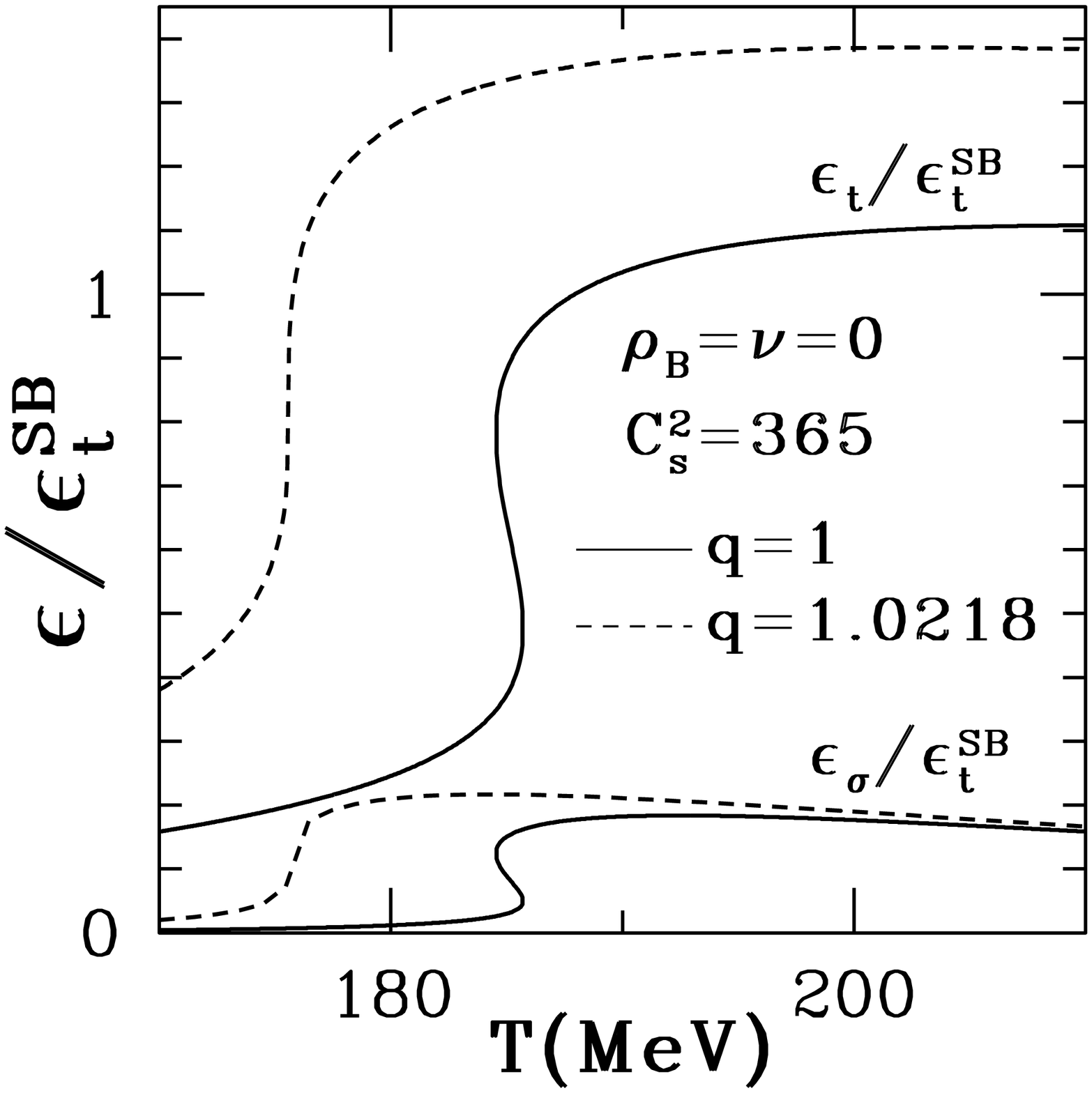,width=3.2truein,height=1.8truein} \hskip .5in}
\caption{The total energy density $\epsilon_t$ and the scalar-field energy
  density $\epsilon_\sigma$ divided by the $q=1$ Stefan-Boltzmann limit
  $\epsilon_t^{SB}$ (given by Eq.~(26) of Ref. \cite{fimPRC2007}) as a
  function of temperature, at zero baryon density of nuclear matter
  ($\gamma=4$). The same value of $C_S^2$ is considered for two different   values of the parameter $q$. 
 In the right panel, we show the same results in the stretched temperature region near the transition point.}
\label{fig2}
\end{figure*}

\begin{figure*}
\centerline{\psfig{figure=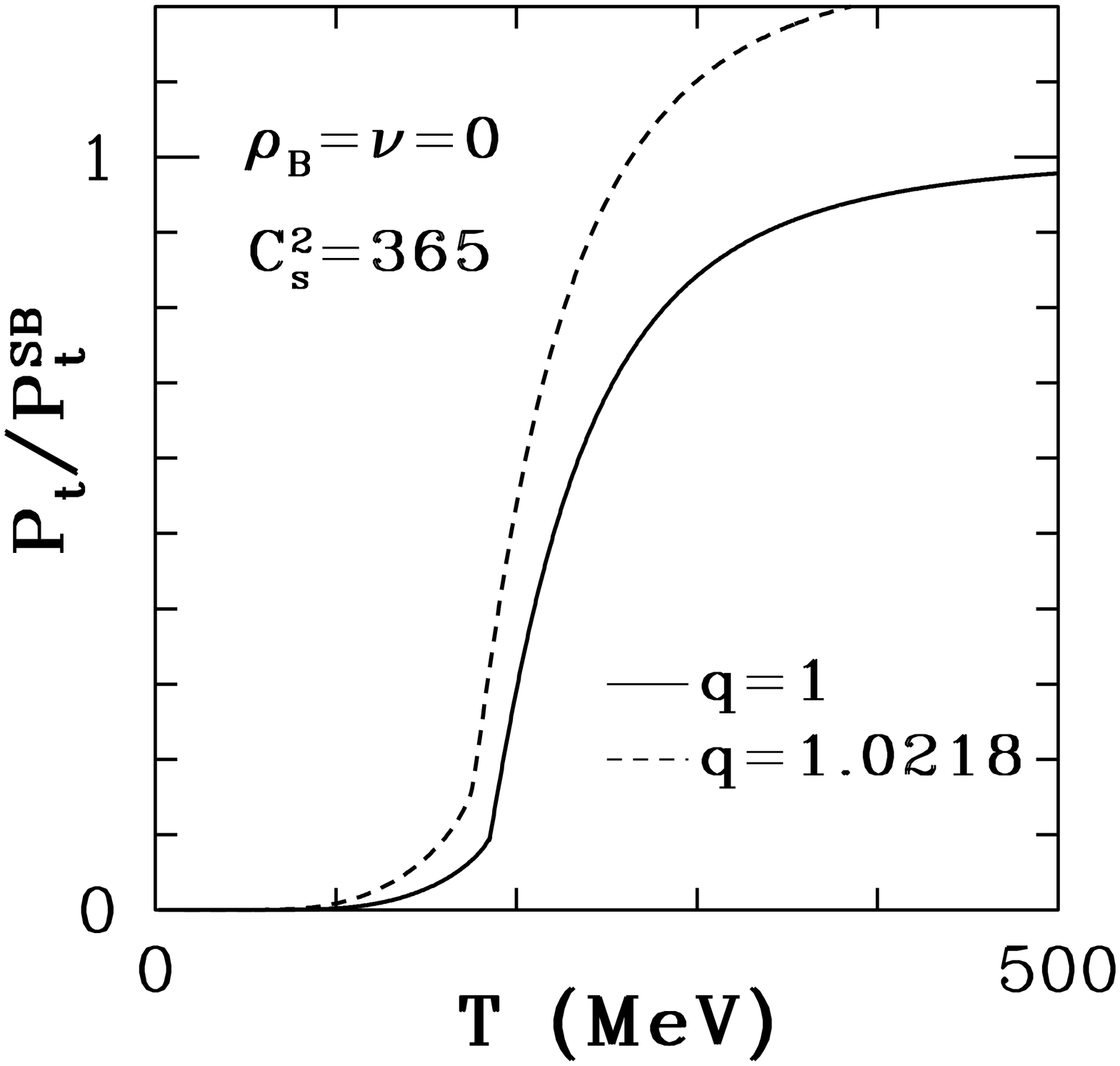,width=3.2truein,height=1.8truein}\hskip
.25in \psfig{figure=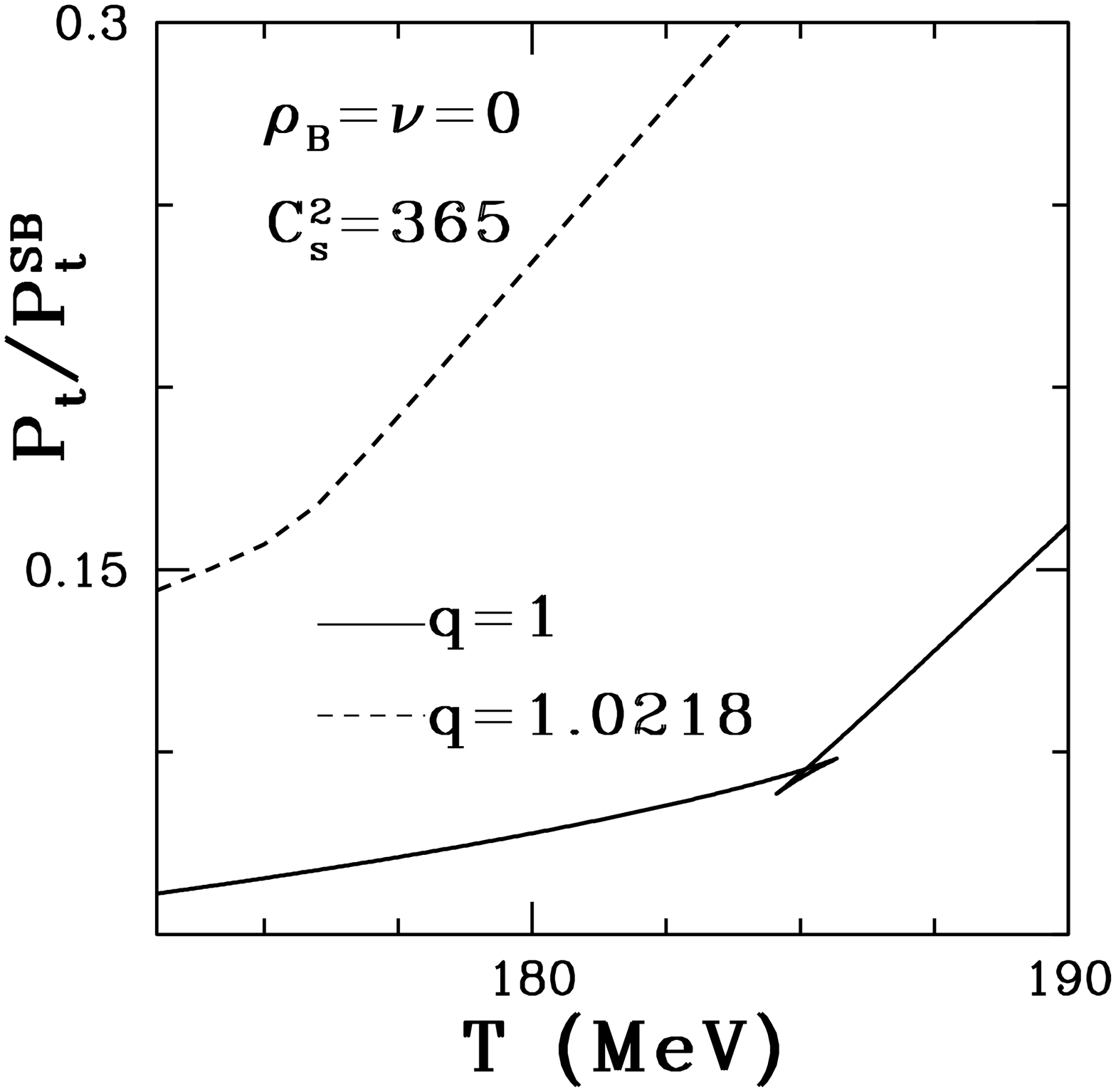,width=3.2truein,height=1.8truein}
\hskip .5in} \caption{The total pressure divided by the corresponding $q=1$
Stefan-Bolztmann limit $P_t^{SB}$ (Eq.~(26) of Ref.~\cite{fimPRC2007}) as
function of temperature. The right panel shows the first order phase transition 
point for $q$ (solid curve) {\bf in the stretched temperature range}. For $q=1.0218$  
(dotted curve) the phase transition is of second order.}
\label{fig3}
\end{figure*}

\begin{figure*}
\centerline{\psfig{figure=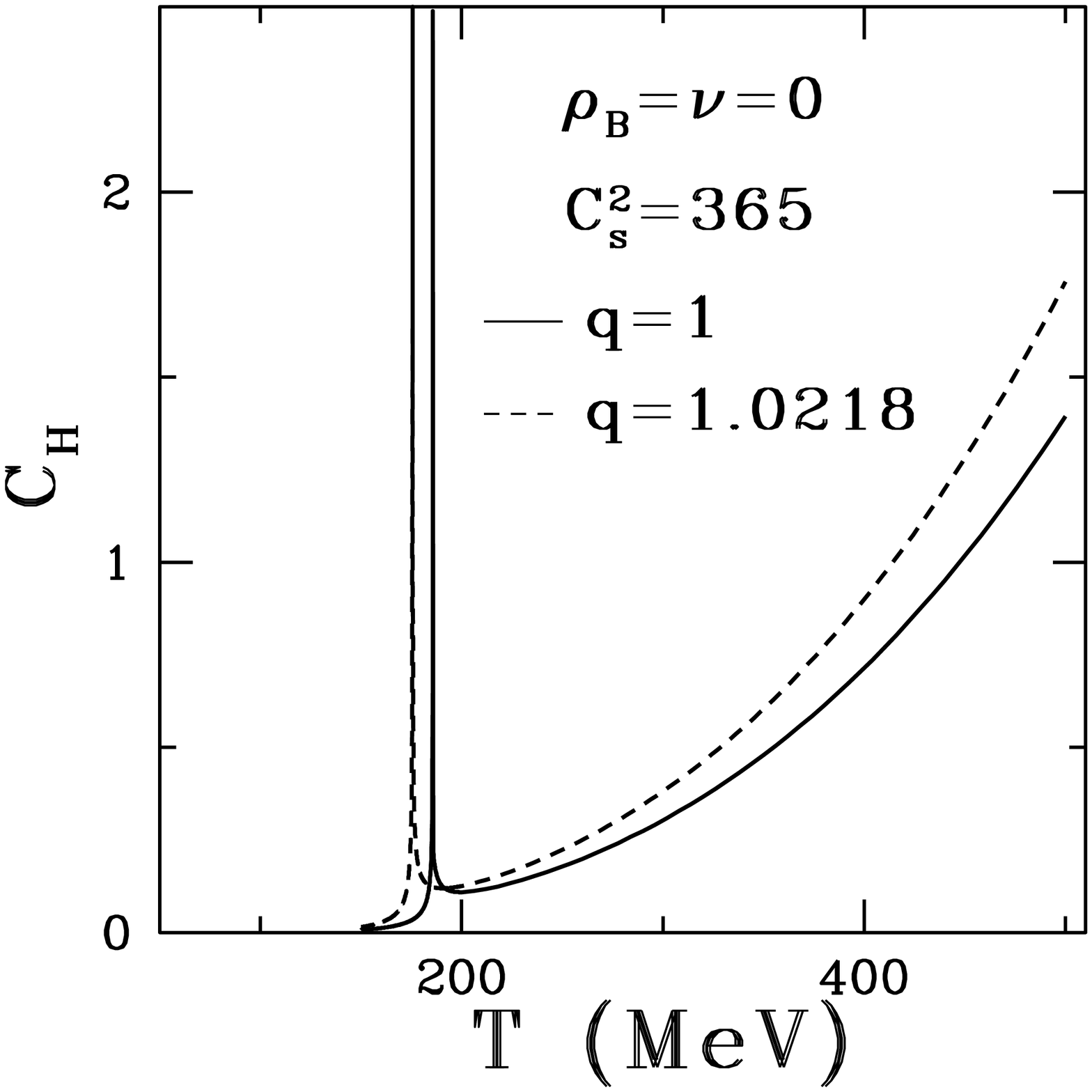,width=3.2truein,height=1.8truein}\hskip.25in
\psfig{figure=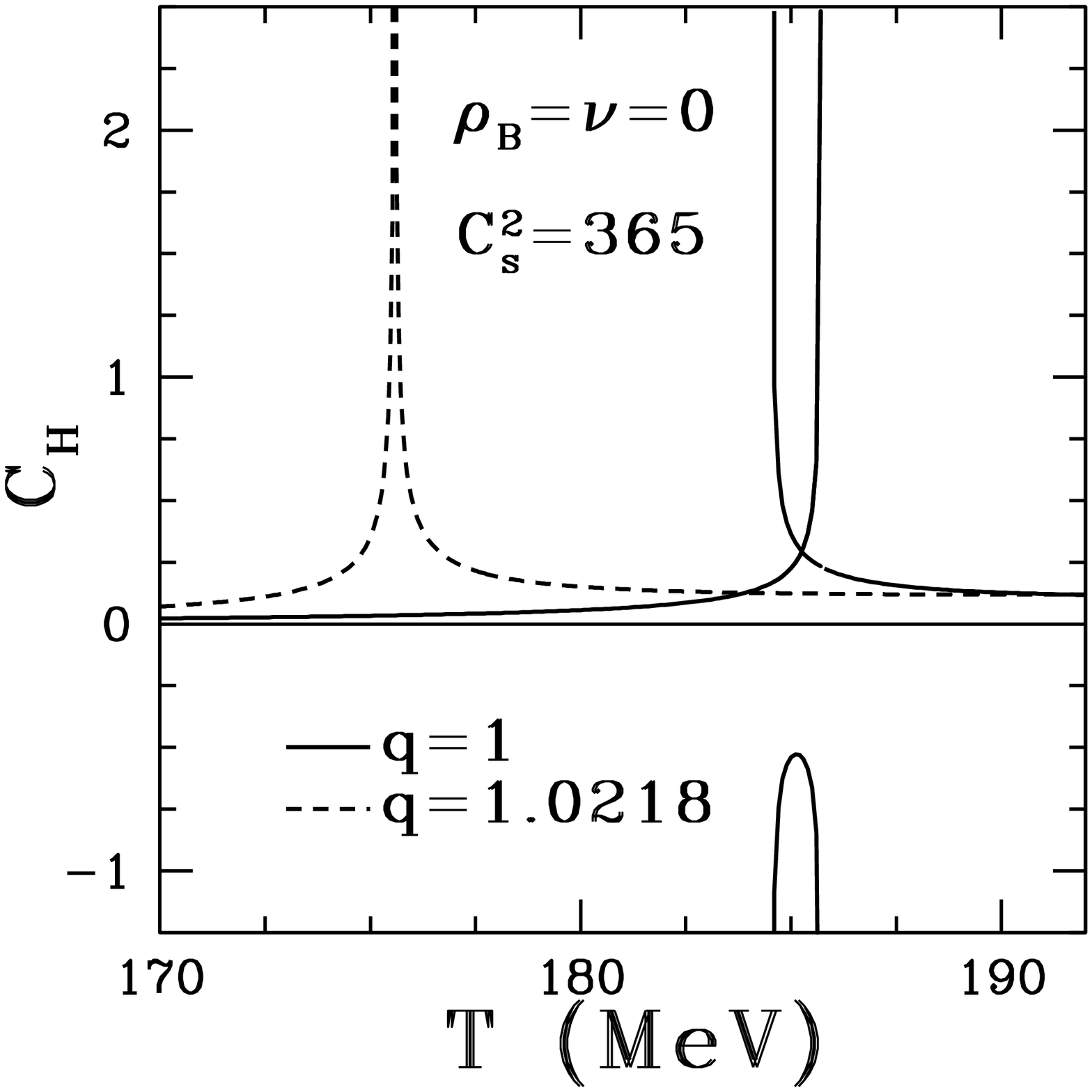,width=3.2truein,height=1.8truein}\hskip .5in}
\caption{The specific heat of nuclear matter ($\gamma=4$) divided by the corresponding $q=1$ Stefan-Boltzmann limit (Eq.~(26) of Ref.~\cite{fimPRC2007}) as function of temperature for the same value of $C_S^2$ and two different values of the parameter $q$. In the right panel, the same results in the stretched region around the phase transition point. For $q=1$ and $q=1.0218$ the phase transitions are, respectively, of the first and second order.}
\label{fig4}
\end{figure*}

\begin{figure*}[tbh]
\centerline{\psfig{figure=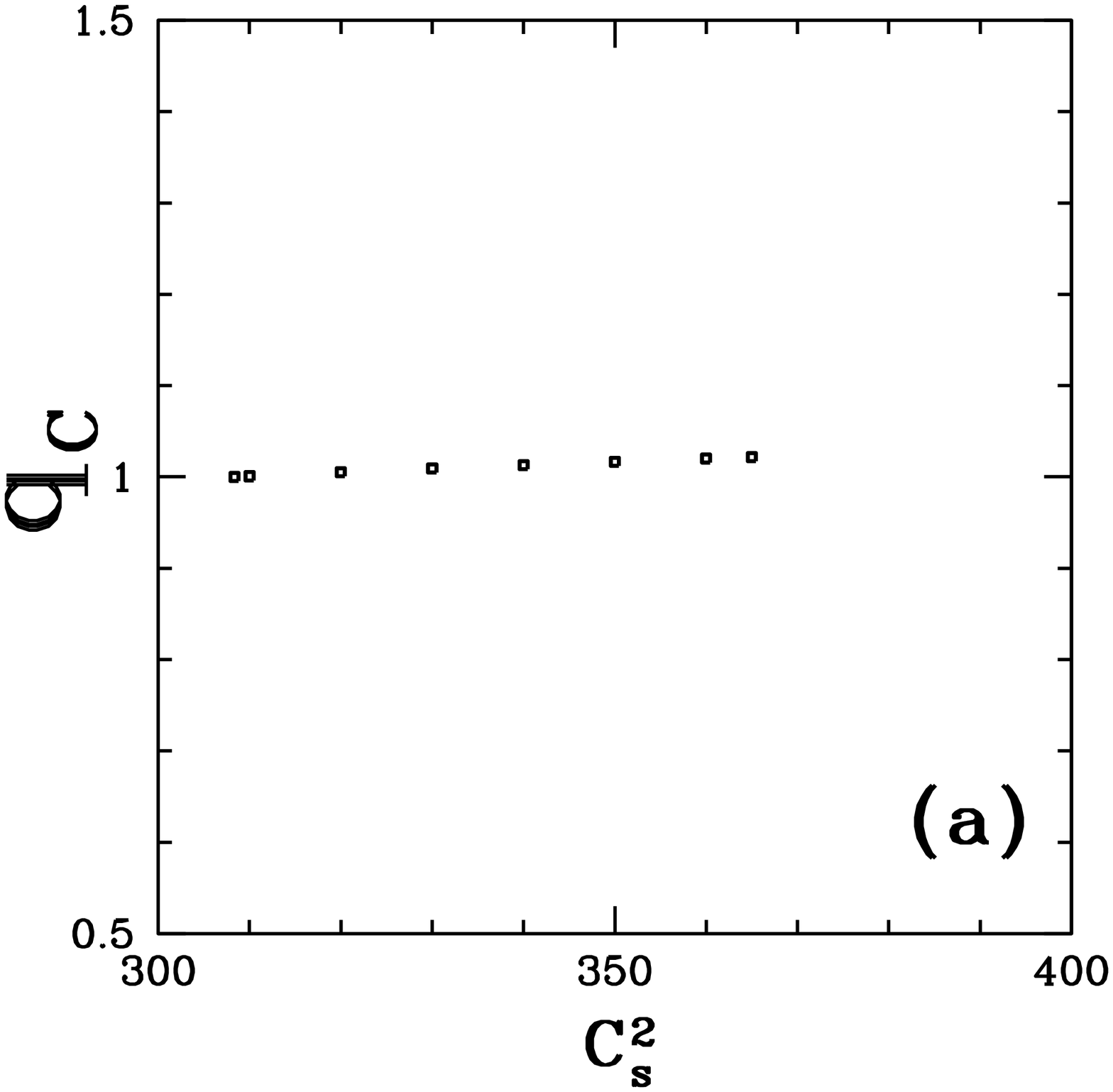,width=2.4truein,height=1.8truein} 
\psfig{figure=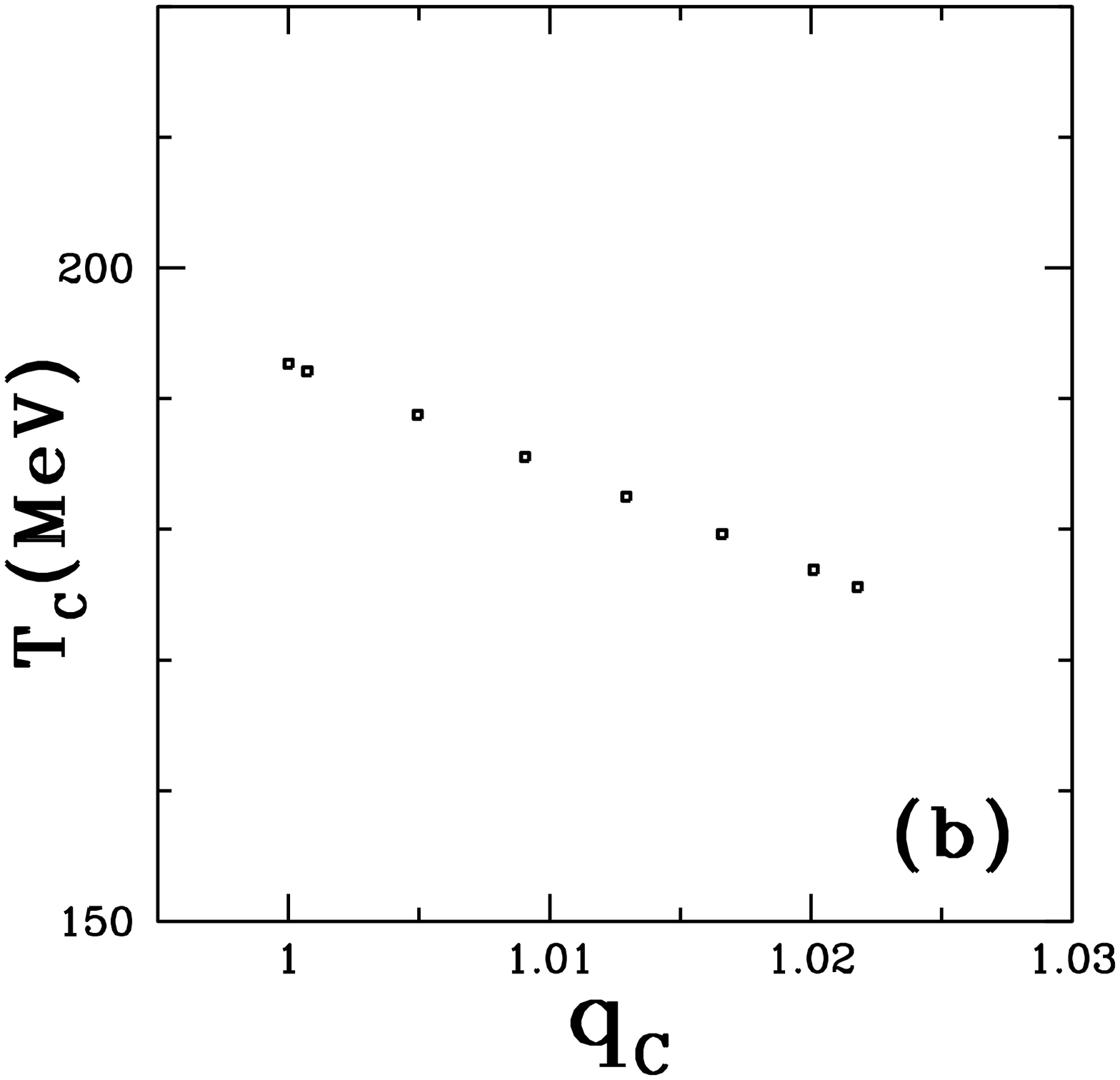,width=2.4truein,height=1.8truein} 
\psfig{figure=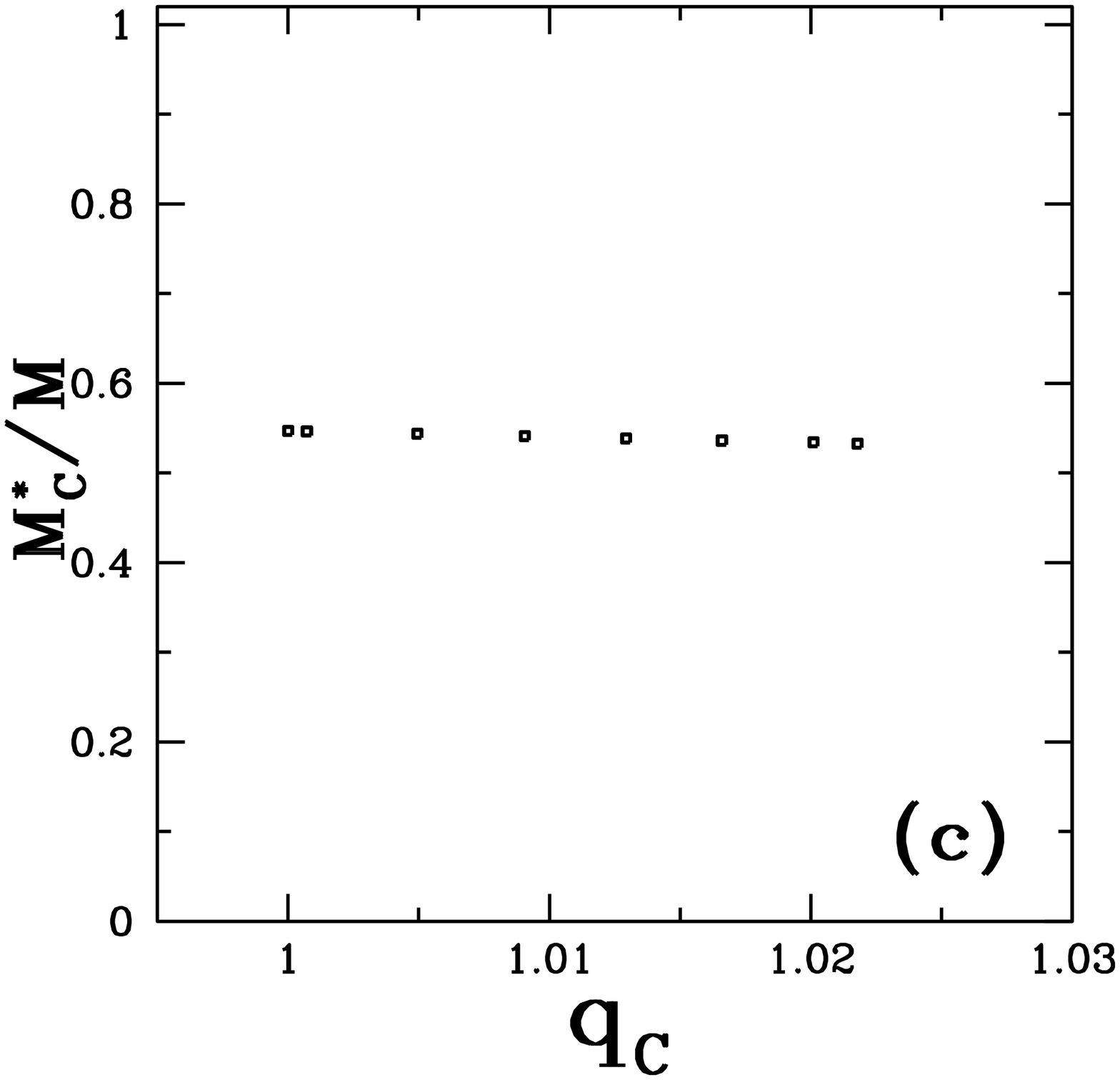,width=2.4truein,height=1.8truein}}
\caption{Panel (a): Values of the critical parameter $q$ for which the phase transitions are of second order as function of the coupling constant $C_S^2$, for nuclear matter ($\gamma=4$) at zero baryon density. Panel (b): The temperature corresponding to Panel (a) as function of $q_C$. Panel (c): The same as in Panel (b) but for the effective mass.}
\label{fig5}
\end{figure*}

Since the function $\tilde{e}_q(x)$ is a \emph{deformed} exponential, the mathematical convergence of the integrals in Eqs. (\ref{RS})-(\ref{press}) must be verified. In Ref.~\cite{fimPRC2007} it was pointed out that, to simultaneously satisfy the convergence
of all integrals in Eqs.(\ref{RS})-(\ref{press}), we have found that $1<q<5/4$, consistent with the limits of $q$ in the range $q\in(0, 2]$ of Ref.~\cite{AR}. Additionally, we also use for the coupling constants the values of reference \cite{SW}, namely\footnote{For the purpose of the present work, the values given in Eq. (\ref{cpcts}) suffices to investigate the effects of the nonextensivity in neutron and nuclear matter.  Variations of the coupling constants, within the acceptable values given in current literature, do not qualitatively affect the conclusions.},
\begin{equation}
\label{cpcts}
\bigg(\frac{g_\sigma}{m_\sigma}\bigg)^2=11.798~{\rm fm^2}~{\rm~and}~
\bigg(\frac{g_\omega}{m_\omega}\bigg)^2=8.653~{\rm fm^2}~,
\end{equation}
which are fixed to give the bind energy $E_{\rm bind}=-15.75$ MeV and
$k_F=1.42$
$\rm{fm}^{-1}$.

\section{the phase structure at zero baryon density}\label{sec6}

In this Section, by considering the same arguments of Ref.~\cite{TSP}, we explore the non-additive phase structure of the effective Lagrangian at vanishing chemical potential and baryon density ($\varrho_B=\nu=0$). To this end, we first consider the range of values for the coupling constant $C_S^2$ given by 
\begin{equation}\label{Cs2}
C_S^2=(\frac{g_\sigma}{m_\sigma})^2 M^2\;
\end{equation}
in the coupling-constant plane shown in Fig. (1) of  Ref.~\cite{TSP}. At vanishing chemical potential, the terms with the baryon density do not appear in Eqs. (\ref{edens}) and (\ref{press}). In what follows, we will take as an example $C_S^2=365$.

Theis {\it{et al.}}~\cite{TSP} showed that the order of phase transition is strongly dependent on the actual value of the coupling constant $C_S^2$. However, in Sec. V of Ref. \cite{fimPRC2007}, it is shown that we can avoid Maxwell construction by the variation of the parameter $q$.  Since the order of transition depends on $C_S^2$ and $q$, the natural question is whether there exists some relation between $C_S^2$ and $q$ at zero baryon density.

In Fig. (1), the sudden drop in $M^*$ around $T\sim185$ MeV determines the abrupt rise of the energy density and pressure.  We note that for $q=1$ (Fermi-Dirac statistics) the self-consistency equation (\ref{ms}) has three solutions around $T\sim185$ MeV imposing a sudden rise in the energy density and a peak in the  specific heat. This behavior is shown in  Figs.~(2)-(4), where the temperature dependence of total energy density, pressure and  specific heat divided by the corresponding high Stephan-Boltzmann temperature limit (for $q=1$), given by Eq.~(26) of Ref.~\cite{fimPRC2007}, are displayed. The behavior of the curves for $q=1$ characterizes a first order phase transition with the pressure curve crossing itself twice at $T\sim185$ MeV. We observe that in this region the energy density and presure are also triple valued and that the specific heat is negative. The value of $q=1.0218$ was obtained by requiring that energy density and pressure to be single valued characterizing a second order phase transition with non-negative specific heat.  This allows to obtain a relation between $C_S^2$ and $q$ explained below~\footnote{We remark that near $T\sim180$ MeV, the effective mass being very small, the system decouples to an almost free-massles nucleon gas. In the present case, this depends on the actual values of $C_s^2$ and $q$. As pointed out in Ref.\cite{TSP}, this is analogous to the expected chiral phase transition in high-temperature QCD.  However, in Walecka theory, there is no liberation of the internal constituents of the nucleons. Thus, the phase transition in QHD theory can not be interpreted as a baryon-quark matter one.}.

\subsection{$C_S^2 - {q}$ relation}

In order to obtain such a relation, in what follows we investigate the thermodynamical behavior of the nuclear matter for several values of the coupling constant $C_S^2$ in the $q \times C_S^2$ plane. For each value of  $C_S^2$, the corresponding parameter $q$, for which the transition is of second order, is determined. Thus, a curve in the $q \times C_S^2$ plane is obtained as shown in Fig. (5a). Below this curve the phase transitions are of the first order and above it the thermodynamical behavior is smooth. Let us now show how the calculation is made via specific heat.

Differently from the treatment discussed in  Ref.~\cite{TSP}, the mathematical structure of the self-consistency equation in our approach is not simple, so that the calculation must be done numerically. We observe that the specific heat calculated from Eq.~(\ref{edens}) is linear in $dM^*/dT$. Thus, whenever there is a sudden fall in $M^*$, we see a peak in the specific heat. By writing 
\begin{equation}\label{Ch}
C_H=\frac{d\varepsilon}{dT}=\frac{d\varepsilon}{dM^*}\frac{dM^*}{dT}
\end{equation}
we can see from Eq. (\ref{ms}) that
\begin{equation}\label{dMdT1}
\frac{dM*}{dT}=\frac{-2C_{M^*}\frac{M^*}{T}\int_0^\infty\frac{2k^2+M^{*2}}{E^*(k)}\;
n_q dk}
{1+2C_{M^*}\int_0^\infty\frac{ k^2-M^{*2}}{E^*(k)} n_q dk}\;
\end{equation}
where
$$C_{M^*}=(g_\sigma/m_\sigma)^2\gamma_N/\pi^2\equiv C_S^2/M^2.$$ 
The singularities of $dM^*/dT$ lie in the curve determined by the
vanishing of the
denominator. Using Eq.~(\ref{ms}), this condition becomes
\begin{equation}\label{dMdT2}
M-2C_{M^*}M^{*3}\int_0^\infty n_q\;\frac{dk}{E^*(k)}=0\;.
\end{equation}
Note that, for $q=1$, we fully recover Eq.~(18) of Ref.~\cite{TSP}.

The number of intersections of the solutions obtained from Eq.~(\ref{dMdT2}) and the self-consistency equation given by Eq.~(\ref{ms}) determines how decoupling happens. We have a first or a second order phase transition, respectively, for two or one intersections. If there is no intersections, the thermodynamical behavior is continuous. The numerical results shown in Fig. (5a) can be summarized as
follows:

\begin{itemize}

\item For $q$ lying below the $q\times C_S^2$ curve the phase transitions are of the first order.

\item For $q$ lying on the $q\times C_S^2$ curve the phase transitions are of the second order. The corresponding values of temperatures and effective masses are shown in Figs. (5b) and (5c). 

\item For $q$ lying above the $q\times C_S^2$ curve the decoupling is
continuous.

\end{itemize}
Therefore, the order of phase transitions depends not only on the actual values of $C_S^2$ but also on the values of $q$.

\section{final remarks}

We have investigated some effects of the Tsallis NESM on the mean field theory of Walecka (QHD) \cite{SW}. Instead of the standard Fermi-Dirac nucleon and antinucleon distribution functions, we have used the $q$-quantum distribution obtained in Ref. \cite{TPM}.  Although there is no long range interactions in nuclear matter (Coulombian interaction is not included in our analysis), it is worth emphasizing that the NESM effects appear due to the strong statistical correlations between elements of nuclear system.

We have examined the phase structure of nuclear matter at high temperature and at zero baryon density. 
The effective Lagrangian of QHD theory is considered for the same set of values of the coupling 
constants in the coupling constant-plane of Ref.~\cite{TSP}. Given that, at vanishing baryon 
density, the term with the baryon density do not appear in the equation of state, a relation between the 
coupling constant $C_S^2$ and the parameter $q$ is obtained. As discussed, such a relation 
determines, in the $q\times C_S^2$ plane, regions of different thermodynamical behaviors. 


{\bf Acknowledgments:} The authors thank CNPq - Brazil for the grants under which this work was carried out.

\end{document}